# Influence of Steel Alloying Components on Martensite Tetragonality

L.S.Kremnev

*Moscow State Technological University «Stankin», Moscow, Russia*
kremnevls@yandex.ru

Abstract: A formerly developed method aimed at determining the constituents of a merged X-ray pattern doublet was applied to estimate the fact that low-carbon martensite with high concentration of strong carbide-forming elements possesses abnormally small tetragonality ($c/a < 1$). The article treats the nature of carbide-forming and non-carbide-forming steel components influence on the tetragonality of martensite.

Martensite tetragonality is known to depend on its lattice spacings "c" – to -"a" ratio [1]. It has been found [1,2] that carbon steel martensite has $c/a > 1$, with this value being directly proportional to the carbon concentration in martensite. This conclusion has found multiple confirmations in a number of researches, but only few of them were devoted to the study of steel alloying components influence on low-carbon martensite tetragonality. It can be explained by the fact that at low carbon concentrations in polycrystalline steel martensite (no more than 0.5 – 0.6% ) the components of its X-ray pattern doublet are merged. This circumstance eliminated the possibility of reliable determining martensite lattice spacings and, hence, its tetragonality. Reference [3] provides a method for martensite doublet de-mergering into its constituent components. The method is based on the assumption that a merged doublet x-ray pattern is not symmetrical about the vertical axis, the one passing through its gravity center. The method developed allows for obtaining the profile of the doublet constituents' pattern lines based on obtaining the difference between the two areas located on either side of the doublet vertical axis. It has been proved [3] that if each doublet component ordinates' sum coincides with the doublet pattern line ordinate (fig.1,3,4,6), the de-mergering performed is the only possible one. Violation of the abovementioned condition witnesses either of the fact that martensite diffraction maximum is overlapped by those of other steel phases' lattice plains, or of biphasic breakdown of martensite, or of the fact that the latter has different crystalline lattice, e.g. the orthorhombic one. Another characteristic feature of the method concerned consists in providing the ability to determine the maximum half-width to be used in calculating microstresses and mosaic block sizes.

The research was carried out using the specimens made of P18 high speed steel Ø 25 mm rolled stock and those of MP18 steel forged 15x15 mm square bars. Composition of steel MP18 is similar to that of steel P18 solid solution after quenching.

Both steels chemical compositions are given in Table 1.

Table 1

Chemical compositions of steels P18 and MP18

| № | Steel grade | Element content, % mass. | | | | | | | |
|---|---|---|---|---|---|---|---|---|---|
| | | C | W | Cr | V | Mn | Si | S | P |
| 1 | P18 | 0.71 | 18.22 | 4.12 | 1.21 | 0.32 | 0.45 | 0.03 | 0.03 |
| 2 | MP18 | 0.46 | 7.12 | 3.82 | 0.61 | 0.40 | 0.38 | 0.02 | 0.02 |

After steel P18 and MP18 heating to 1280 $^{o}$C and 1290 $^{o}$C respectively, 8 mm-high annealed specimens were quenched in a chlorine-barium bath. Before heating the specimens the bath had been deoxidized with magnesium fluoride. The period of specimens' hold in the bath equals 6 minutes, quenching the P18 specimens was performed in oil medium and those of MP18 were quenched in 10% water solution of NaCl in order to prevent carbide precipitation from austenite in process of quenching. Specimens tempering took place in potassium nitrate – based molten bath within 350 – 625 $^{o}$C interval, with their hold time being 1 hour (after heating the specimens up to the prescribed temperature). After their heat-treatment and flushing the butts the specimens were ground with intensive simultaneous cooling and electrolytically polished in 5% hydrochloric acid water solution at current density of 0.15 A/cm2 till 0.10 mm - thick surface layer' removal. Electrolyte temperature was no higher than 400$^{o}$ C. X-ray research of the samples was carried out at X-ray diffractometer УРС – 50ИМ in FeK$_{\alpha}$- radiation with β –filter. Carbon concentration in the investigated steel martensite was determined by its {110} and {211} diffraction maximums. Error of carbon concentration measurement in martensite didn't exceed 15% and 10% for quenched and tempered steels respectively [3]. In the figures provided the last character after the point is to be considered indubitable.

Fig.1 demonstrates X-ray patterns of P18 steel in the condition after 1100 and 1280 $^{o}$C quenching. One can easily see that carbon concentration in P18 steel determined by {110} peak after quenching from 1100 $^{o}$C and 1280 $^{o}$C equals 0.30 and 0.42% respectively. These results coincide with the data of [4, 5, 6 (Fig. 2)] researches, where carbon concentration in steel P18 solid solution after quenching was determined by means of phase chemical analysis.

On the X-ray patterns presented in Fig.1 one can see that angle θ of (101)(011) martensite planes diffraction peaks are less than angle θ of (110) maximum. It means that (101)(011) interplanar spacings exceed those of (110) and, hence, martensite tetragonality of quenched steel P18 with 0.30% C and 0.42% C is greater than unity (c/a =1.014 and 1.019 respectively)

Results of determining carbon concentration in the martensite of steel P18 as a function of its tempering temperature are presented in Table 2. Fig. 3 provides the examples of X-ray technique application in determining carbon concentration in P18 martensite after tempering at 350 $^{o}$C and 560 $^{o}$C (triple hold, each one - one-hour long).

Table 2

Action of quenching and tempering temperature on carbon concentration in P18 and MP18 martensite.

| Designation | Carbon concentration in martensite, % (mass.) | | | | | |
|---|---|---|---|---|---|---|
| | After quench | After tempering at temperatures, $^{o}$C | | | | |
| P18 | 1280 $^{0}$C | 350 | 400 | 560 | 600 | 625 |
| | 0.42 | 0.24 | 0.27 | 0.22 | 0.26 | <0.1 |
| | 1100 $^{0}$C | | | | | |
| | 0.30 | | | | | |
| MP18 | 1290 $^{0}$C | | | | | |
| | 0.43 | | | 0.17 | | |

Attention should be drawn to the fact that interplanar spacing of (211) (121) planes of tempered martensite lattice is greater than that of (112) planes regardless of actual tempering temperature within the studied temperature range. It can be easily demonstrated* that in this case tetragonality of tempered P18 steel martensite is less than unity (c/a = 0.989 and 0.990 after 350$^{0}$C and 560 $^{0}$C tempering respectively). A similar result was obtained in studying high-speed steel P6M5.

To validate these conclusions and to completely eliminate the influence of residual austenite weak lines and those of the carbides which hadn't dissolved in P18 steel when heated to 1280 °C MP18 steel specimens were subjected to x-ray analysis (Fig.4). It was predetermined that after-quenching structure of this steel doesn't contain either residual austenite or undissolved carbides (in accordance with x-ray and metallographic analysis results). The latter can be additionally confirmed by the fact that carbon concentration in the martensite of quenched steel MP18 appeared to equal 0.43% (Fig.4a), that is, it is similar to that in the chemical composition of this steel (Table 1). Alongside with these, it can be seen, that interplanar spacings (211)(121) of quenched MP18 steel martensite lattice are less than those of (112) plates, whereas similar interplanar spacings of tempered martensite have been changed into the inverse ratio (Fig.4b). Thus, martensite tetragonality of quenched MP18 steel $c/a > 1$, whereas tetragonality of the same steel after tempering is less than unity ($c/a = 0.992$). This result confirms the conclusions obtained using the specimens of P18 (Fig. 3) and P6M5 steels.

Hence, in process of tempering the samples of quenched P18, P6M5 and MP18 steels the following transformations take place in these steels martensite:
1. Precipitation of of carbon atoms from martensite lattice
2. Formation of martensite with abnormally low tetragonality $c/a<1$ (Fig.5).

Hence, the latter transformation which isn't observed during carbon and low-alloyed steels tempering is characteristic of high-alloyed steels, namely, of high-speed and some die tool steels. Carbon atoms, as is known from [2], can occupy octahedral interstitial lattice spaces in any of their three sublattices. So one can suppose that the observed tetragonality change is caused by carbon atoms transfer from one octahedral interstitial space sublattice to another one in process of high alloyed steels tempering. Concurrently, constant «c» of martensite lattice cell turns into «a» constant, and vice versa [2].

The testimony of this transformation can be found in the results of «c» and «a» martensite lattice constants comparison in MP18 steel with 0.17% C (Table.2) and ~ 6.5% (mass) W after tempering at 560 °C (by data of Fig.2), which has $c/a < 1$, and the martensite of a plain steel with 0.17% C ($c/a > 1$).

Table 3
Martensite lattice constants «c» and «a» [nm] of non-alloyed steel 0.17% C in martensite ($c/a > 1$) and steel MP18 after tempering 560 °C ($c/a < 1$).

|   | Constant «c» | Constant «a» | c/a | a/c |
|---|---|---|---|---|
| 1. Martensite of steel MP18 | 0.2864 | 0.2887 | 0.992 | 1.008 |
| 2. Martensite of steel with 0.17% C** | 0.2886 | 0.2863 |  | 1.008 |

( **Calculation was performed by means of correlations [2] established between «c», «a» values and carbon concentration in plain steel martensite.)

One can see that the value of MP18 steel constant «c» is practically equal to constant «a» in 0.17% C carbon steel martensite, and constant «a» of the first one – to constant «c» of the second one. Thus, in process of tempering MP18 martensite constant «c» transforms into constant «a», and constant «a» - into constant «c». Hence, in this case carbon concentration in the tempered martensite shall be directly proportional to the retio of its «a» and «c» lattice constant. In point of fact it turned out that the results of determining carbon concentration in the martensite of tempered steels P18 (Table 2) calculated on the basis of this assumption coinside with those off phase chemical analysis (Fig.2 [6]).

Comparison of Fig. 4a and Fig.4b confirms the supposition consisting in the fact that carbon concentration decrease from 0.43% to ~ 0.20% in the martensite of quenched steel MP18 during its tempering within the temperature range of 350 – 560 °C, which arouses carbon atoms transfer from one sublattice of octahedral interstitial spaces to another one. At the same time, tungsten atoms aren't involved in this martensite transformation because its concentration in the martensite stays the same ( Fig.2).

The driving force of this transformation is referred to as the tendency of martensite crystalline lattice to its minimal elastic energy condition [2]. It is evident that with carbon concentration reduction during tempering and the decrease in tetragonality the distance reduces between the tungsten and carbon atoms located along the <[100]> martensite lattice directions.

Apparently, carbon and tungsten atoms contingence, the latter being a strong carbide-forming element, leads to their interaction. This circumstance additionally reduces martensite lattice constant «c». As a result interplanar spacings of (101) (011) planes reduce and these are the ones between which carbon atoms are located. This in turn causes lattice elastic energy increase. Probably, this is the reason for carbon atom transfer into the adjacent octahedral interstitial site, characterized by the absence of tungsten atoms in its proximity, and the lattice elastic energy reduces. As a result carbon atom is placed between (110) planes of martensite lattice, in this case ($c/a < 1$) the distance between them being larger than that between (101)(011) planes. These shall be followed by the cell volume increase. Contents of Table 2 allow for martensite lattice cell volumes comparison of 0.17% carbon steel ($V_1 = c \times a^2$) and steel MP18 ($V_2 = a^2 \times c$). One can get convinced that $V_2 > V_1$ by ~ 0.9%.

It is remarkable that if carbon and tungsten concentrations in martensite (steel P18 after quenching from 1100 °C, Fig. 1a) are not high, or the contents of these components in martensite is considerable ( steels P18 and MP18 after quenching from 1280 °C and 1290 °C respectively, Fig.2a and Fig..4a) abnormal martensite tetragonality is absent.

In low-carbon (~ 0.2% C(mass.)) high-alloyed martensite (~7% W(mass)) of tempered steels carbon and tungsten concentration comprise ~1%(at.) and ~ 2%(at.) respectively. Hence, 1 carbon and 2 tungsten atoms fall within every 50 martensite lattice cells. Apparently, there seem to exist certain concentrations of carbon and carbide-forming components capable of causing decrease of tetragonality up to $c/a < 1$ in some lattice cells, and their number appears to be enough to rearrange the entire martensite lattice. To solve this problem additional researches have to be conducted.

In Fig.6 one can see X-ray pattern of a high-alloyed hot-die steel specimen (grade 4Х5В2ФС, GOST 5950 – 2000) after quenching from 1100 °C. Carbon concentration in the martenssite of this steel after quenching is 0.26% with its tetragonality $c/a < 1$ equaling 0.988.Thus, it isn't only after tempering but after quenching as well that low-carbon martensite with high concentration of carbide-forming elements has tetragonality $c/a < 1$.

In light of the problem discussed the question naturally arises regarding non-carbide-forming alloying components influence on martensite tetragonality. Monograph [2] provides some previously reported investigation results concerning the study of aluminum, nickel and manganese influence on martensite tetragonality. In studies [7 – 10] it has been found that under aluminum and nickel influence steel martensite gains abnormally high tetragonality. Yet, the papers contain no explanations for this anomaly. Attention should be drawn to the fact that neither nickel nor aluminum is referred to as carbide-forming components. In iron-base alloys they often behave as graphitizing components. Therefore their influence on martensite tetragonality shall be opposite to that of strong carbide-forming elements – tungsten, molybdenum and others.

So-called carbide-forming elements are known to reduce carbon chemical potential (activity factor) in iron solutions whereas non-carbide-forming ones increase this potential.

In reference [11] it has been shown, that carbon atoms distribution over different octahedral interstitial spaces is determined by the energy of elastic interaction between carbon atoms. For this reason one should take into consideration considerable and opposing influence of alloying components (either increasing or decreasing carbon activity in iron) on martensite tetragonality.

Conclusions:
1. Tetragonality of high-alloyed steel martensite doesn't only depend on carbon concentration; it is also affected by the content of alloying components and the peculiarities of their interaction with carbon atoms.
2. The presence of sufficient amounts of carbide-forming components in the lattice of low-carbon martensite reducing carbon activity factor can cause its tetragonality reduction up to $c/a < 1$.
3. The presence of sufficient amounts of non-carbide-forming components increasing carbon activity factor leads to abnormally high martensite tetragonality.
4. In process of P18 and P6M5 high speed steels tempering martensitic transformation takes place which is connected with reciprocal transfer of «c» and «a» constants of its crystalline lattice.*
5. Carbon concentration in the martensite with its tetragonality $c/a < 1$ is proportional to a/c ratio a/c and is determined by their carbon concentration dependence established in reference[1] with their reciprocal transfer taken into consideration. .

---

*For tetragonal system $1/d^2(hkl) = (h^2+k^2)/a^2 + l^2/c^2$, where d – distance between crystalline lattice planes with Miller indexes (hkl). As it comes from the X-ray pattern, low-carbon, low-alloy martensite angle $\theta_2$ of diffraction maximum $d_{(211)(121)}$ is less than the value of $\theta_1$ from planes $d_{(112)}$ - ( $\theta_2 < \theta_1$). Hence, $d_{(211)(121)} > d_{(112)}$, as it follows from Bragg equation; $1/d^2_{(211)(121)} - 1/d^2_{(112)} < 0$, that is $(2^2 + 1^2)/a^2 - 1^2/c^2 - (1^2 + 1^2)/a^2 - 2^2c^2 < 0$  $3/a^2 - 3/c^2 < 0$, $(1/a^2 - 1/c^2) < 0$, $c < a$, $c/a < 1$. It goes without saying that in case $\theta_2 > \theta_1$  $c > a$, $c/a > 1$

**REFERENCES**

1. Seljakov N.Ja, Kurdumov G.V., Gudchov I.T. Journal of Applied Physics, **2,** 51 (1927).
2. G.V. Kurdumov, L.M.Uyevsky, G.I. Entin, *Transformations in Iron and Steel* (Nauka, Moscow, 1977).
3. Kremnev L.S., Adaskin A.M., Bogolubov A.V. Industrial Laboratory. **9,** 1086 (1971).
4. Kupalova I.K. Industrial Laboratory. **1**, 27 (1983).
5. I. Artinger, *Tool steels and heat treatment* (Metallurgia, Moscow, 1982)
6. Yu.A. Geller, *Tool steels* (Metallurgia, Moscow, 1983)
7. Watanabe M., Wayman G.E. Scripta Met. **5,** 109 (1971).
8. Lysak L.I., Artimuk S.A., Polishchuk Yu.M. The Physics of Metals and Metallography, **35**, 098 (1973).
9. Mikhailova L.K. Doklady Akademii Nauk, **216**, 778 (1974).
10. Kurdumov G.V., Mikhailova L.K., Khachaturian A.G. Doklady Akademii Nauk **215,** 578 (1974).
11. Khachaturian A.G. The Physics of Metals and Metallography, **9**, 2861 (1967).

**Figures captions**

Fig.1. X-ray peaks of P18 samples after quenching:
 a) from 1100 $^0$C;  b) from 1280 $^0$C.
   1. Experimentally obtained X-ray peak of merged X-ray patterns (101)(011) – (110) of martensite lattice corrected with the presence of residual austenite taken into consideration.
   2. Calculated X-ray peak of residual austenite (111) plain.
   3. Calculated X-ray peak of martensite (101)(011) plains.
   4. Calculated X-ray peak of martensite (011) plain.

Fig.2. Chemical composition (% $_{mass.}$) of P18 steel solid solution.
     (quenching from 1280 $^0$C) as a function of tempering temperature[3].

Fig.3. X-ray peaks of P18 samples after quenching from 1280 $^0$C [6].
    a) Tempering 350 $^0$C, 1 hour;  b) Tempering 560 $^0$C, 1 hour, 3 times.
   1. Experimentally obtained X-ray peak of merged X-ray patterns (211)(112) – (112) of martensite lattice.
   2. Calculated X-ray peak of martensite (211)(121) plains.
   3. Calculated X-ray peak of martensite (112) plains.

Fig.4. . X-ray peaks of MP18 samples:
      a) quenching 1290 $^0$C; b) quenching 1290 $^0$C, tempering 560 $^0$C, 1 hour, 3 times.
   1. Experimental peak of merged (211)(121)- (112) martensite patterns.
   2. Calculated peak of martensite (211)(121) plains.
   3. Calculated peak of martensite (112) plains.

Fig.5. Rearrangement of highly-alloyed martensite crystalline lattice taking place during P18 and M18 steels tempering.
    a) –tetragonal martensite lattice after quenching (c/a > 1),
    b) – tetragonal martensite lattice after tempering 350 – 600 $^0$C (c/a < 1).

Fig.6. X-ray patterns of 4X5В2ФС steel specimen
    (Quenching 1100 $^0$C).
   1. Experimental X-ray peak of merged X-ray patterns (211)(121) – (112) of martensite lattice.
   2. Calculated peak of martensite (211)(121) plains.
   3. Calculated peak of martensite (112) plains.
   4. Total ordinates of calculated doublet components.

**Figures**

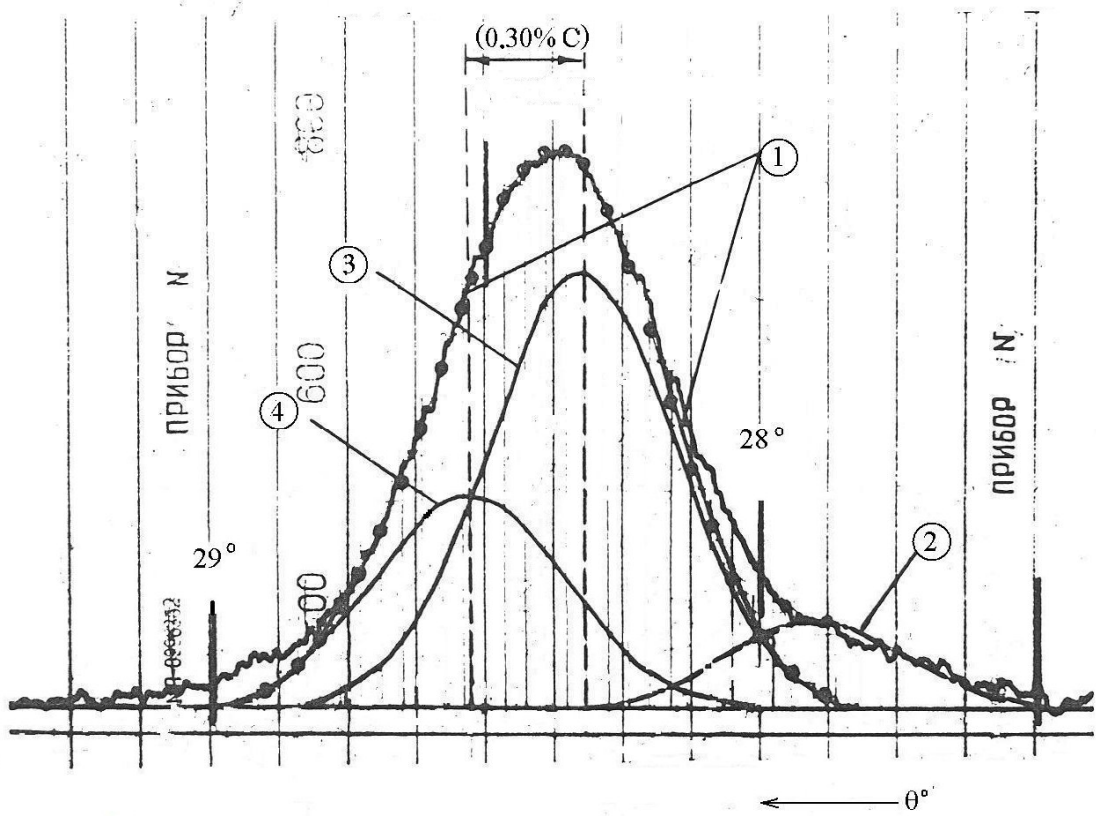

a)

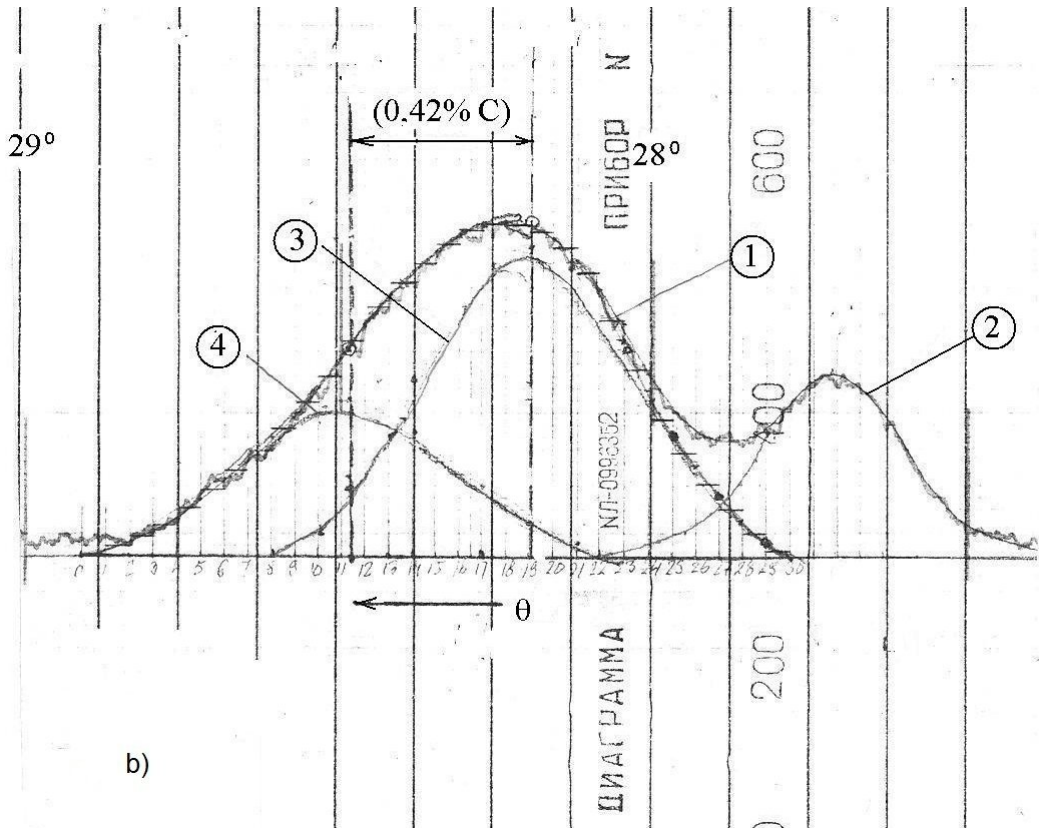

b)

**Fig.1**

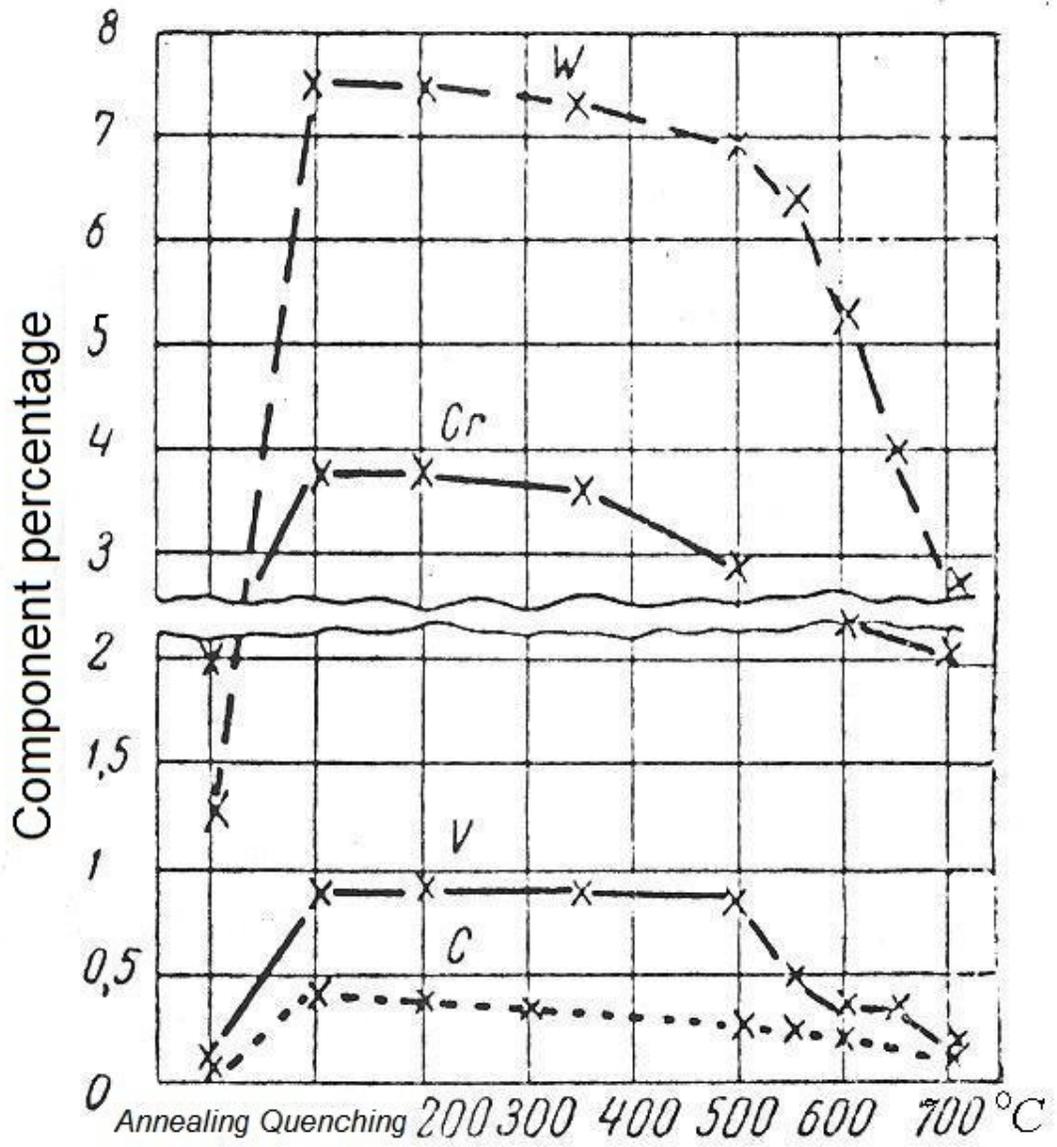

**Fig.2 [6]**

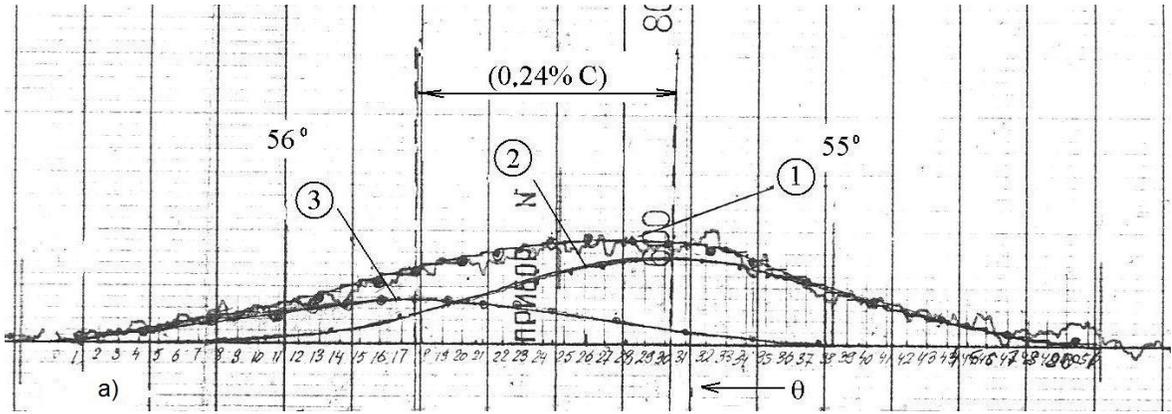

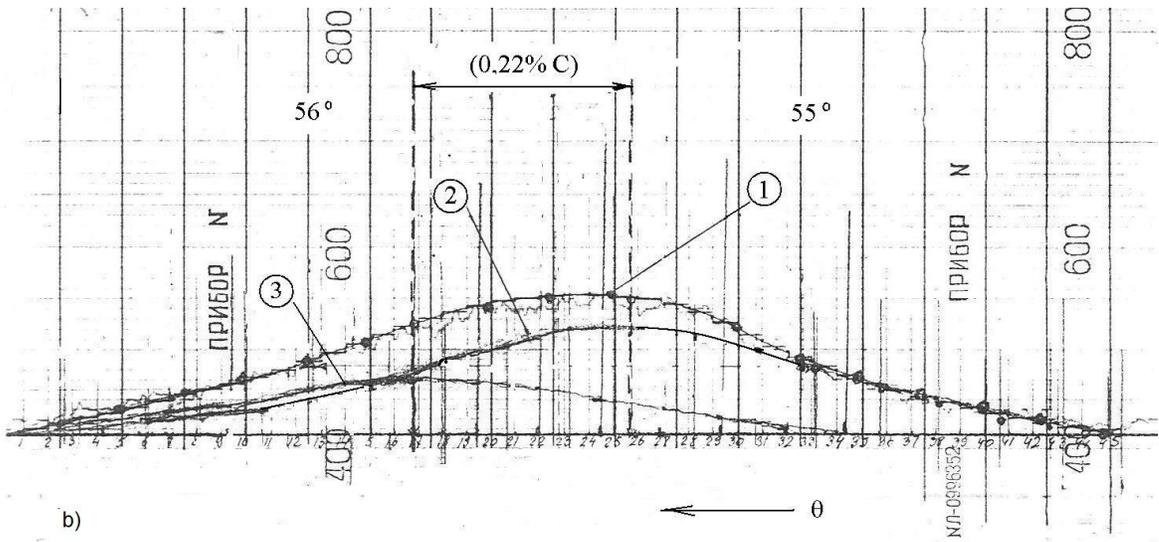

**Fig.3**

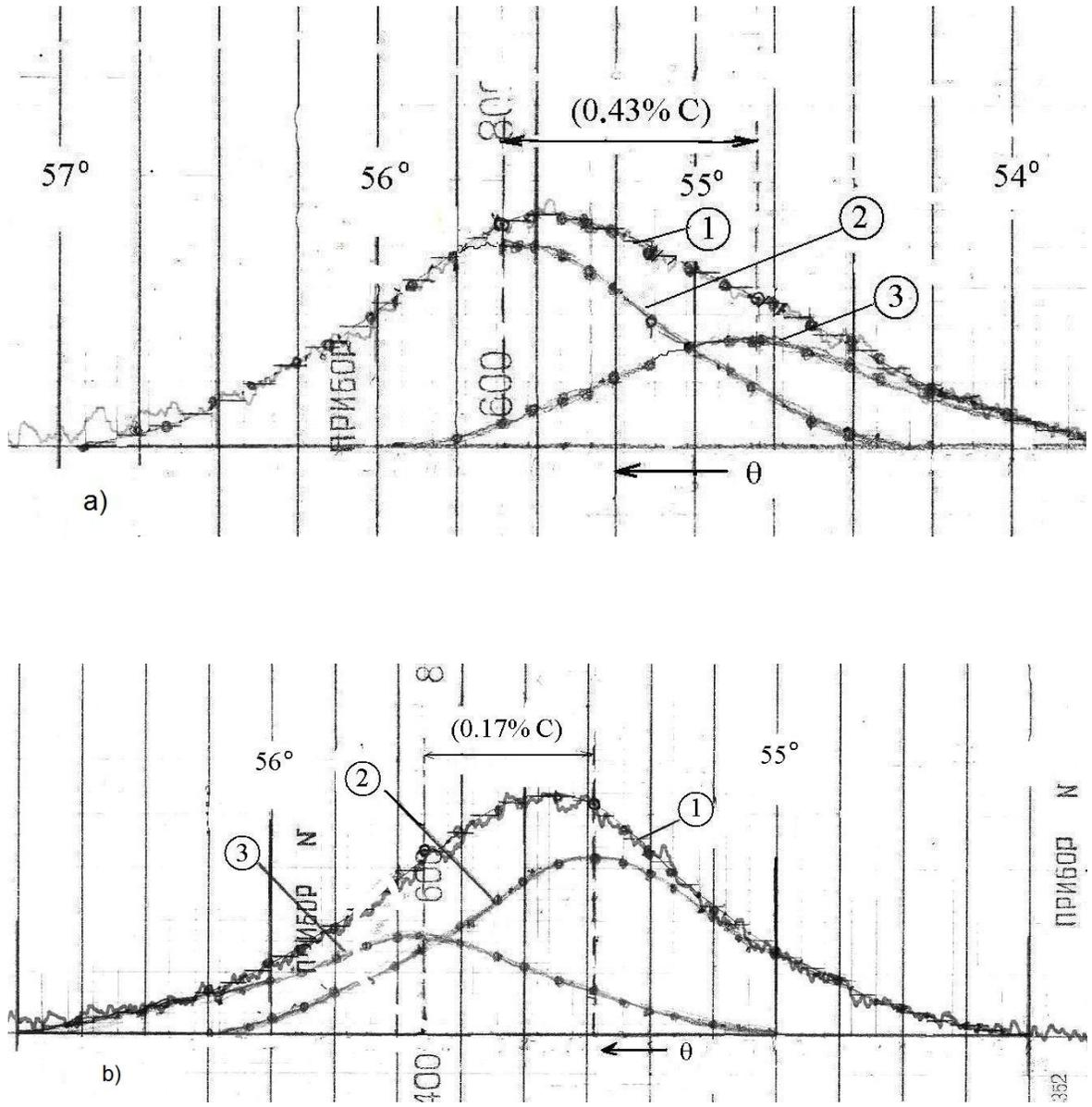

**Fig.4**

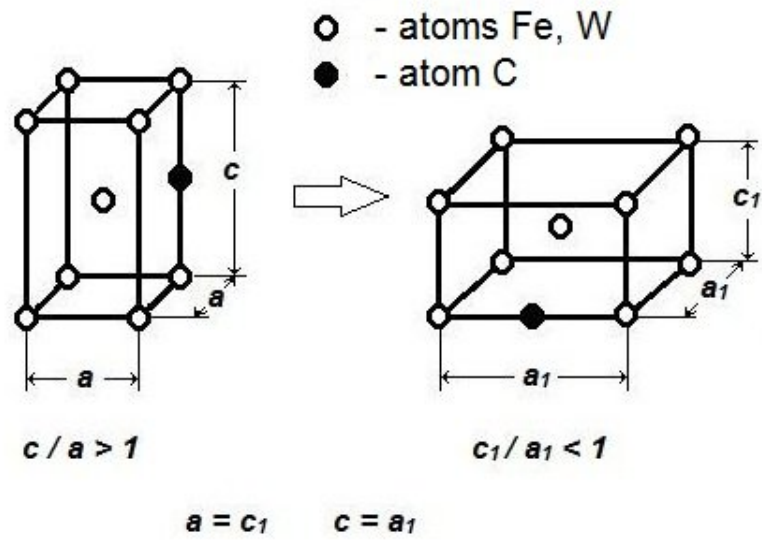

$c/a > 1$          $c_1/a_1 < 1$

$a = c_1$    $c = a_1$

a)          b)

**Fig.5**

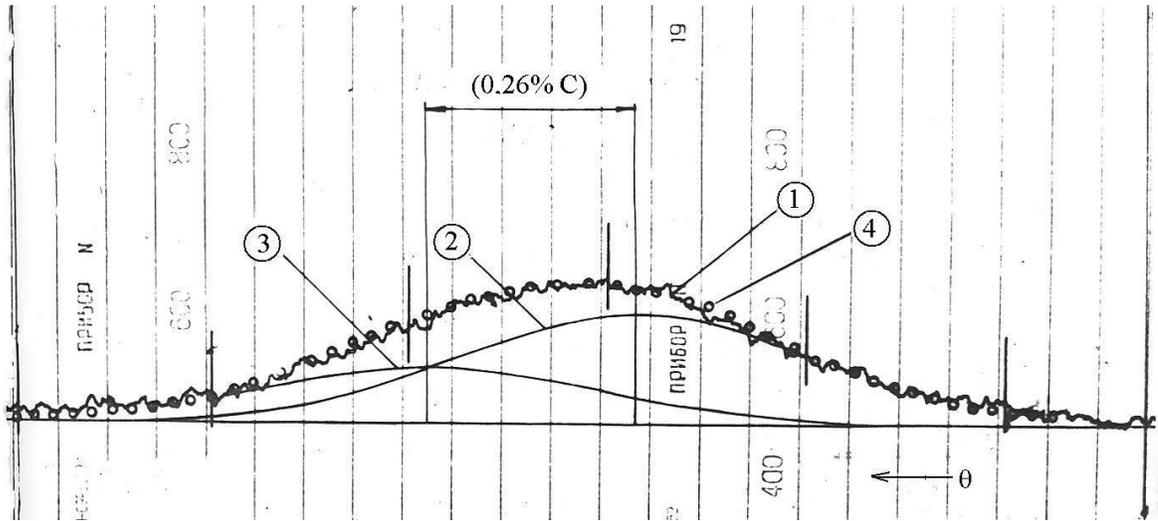

**Fig.6**